\RequirePackage[hyphens]{url} %

\documentclass[sigconf,edbt,dvipsnames]{acmart-edbt2019-arxiv} %

\usepackage{enumitem} %
\setitemize{leftmargin=1.6em, itemsep=0.17em, topsep=0.2em, parsep=0.1em, partopsep=0.1em} %
\setenumerate{leftmargin=1.8em, itemsep=0.21em, topsep=0.2em, parsep=0.1em, partopsep=0.1em}

\usepackage{graphicx}
\usepackage{balance}  %
\usepackage{enumitem}
\usepackage{algpseudocode}
\usepackage{MnSymbol}
\usepackage{color}
\usepackage{epstopdf}
\usepackage{pgfplots}
\usepackage{microtype}
\usepackage{newfloat}
\usepackage{subcaption}
\DeclareFloatingEnvironment[name={Listing}]{lstfloat}

\usepackage{makecell} %
\usepackage{multirow}
\usepackage{extarrows} %

\pgfplotsset{compat=1.13}
\usetikzlibrary{plotmarks}
\usetikzlibrary{patterns}

\definecolor{chartblue}{RGB}{0,163,255} %
\definecolor{chartred}{RGB}{255,0,0} %
\definecolor{chartgreen}{RGB}{47, 227, 129}

\algdef{SE}[DOWHILE]{Do}{DoWhile}{\algorithmicdo}[1]{\algorithmicwhile\ #1}%

\usetikzlibrary{fit}
\newcommand\tikzmark[1]{%
  \tikz[remember picture,overlay]\node[inner xsep=0pt, xshift=0.2em] (#1) {};}
\usepackage{twoopt}
\newcommandtwoopt\BasicBlock[5][2.5cm][2cm]{%
    \begin{tikzpicture}[remember picture,overlay]
      \coordinate (aux) at ([xshift=#1]#4);
      \node[inner ysep=4pt,yshift=0.2ex,draw=black,dotted,%
        fit=(#3) (aux),baseline] 
        (box) {};
      \node[anchor=north east, %
        font=\sffamily\small, align=right, fill=black, text=white,
        inner sep=0.1em]
        at (box.north east) {#5};
    \end{tikzpicture}%
}

\newif\ifdraft

\draftfalse

\ifdraft
  \newcommand{\reviewer}[1]{{\bf\color{orange}Reviewer: #1}}
  \newcommand{\reviewerresolved}[1]{{\bf\color{green}Reviewer: #1}}
  \newcommand{\gabornote}[1]{{\bf\color{green}Gabor: #1}}
\else
  \newcommand{\reviewer}[1]{{}}
  \newcommand{\reviewerresolved}[1]{{}}
  \newcommand{\gabornote}[1]{{}}
\fi

\renewcommand{\baselinestretch}{0.958}

\setcopyright{rightsretained}

\acmDOI{}

\acmISBN{XXX-X-XXXXX-XXX-X}

\acmConference[EDBT 2019]{22nd International Conference on Extending Database Technology (EDBT)}{March 26-29, 2019}{Lisbon, Portugal} 
\acmYear{2019}

\begin{document}

\title{Labyrinth: Compiling Imperative Control Flow to Parallel Dataflows}

\author{G\'abor E. G\'evay}
\affiliation{\institution{Technische Universit\"at Berlin}}

\author{Tilmann Rabl}
\affiliation{\institution{Technische Universit\"at Berlin}}
\affiliation{\institution{DFKI GmbH}}

\author{Sebastian Bre\ss}
\affiliation{\institution{Technische Universit\"at Berlin}}
\affiliation{\institution{DFKI GmbH}}

\author{Lor\'and Madai-Tahy}
\affiliation{\institution{Technische Universit\"at Berlin}}

\author{Volker Markl}
\affiliation{\institution{Technische Universit\"at Berlin}}
\affiliation{\institution{DFKI GmbH}}

\renewcommand{\shortauthors}{G\'abor E. G\'evay et al.}  %

\begin{abstract}

Parallel dataflow systems have become a standard technology for large-scale data analytics. Complex data analysis programs in areas such as machine learning and graph analytics often involve control flow, i.e., iterations and branching. Therefore, systems for advanced analytics should include control flow constructs that are efficient and easy to use. A natural approach is to provide imperative control flow constructs similar to those of mainstream programming languages: while-loops, if-statements, and mutable variables, whose values can change between iteration steps.

However, current parallel dataflow systems execute programs written using imperative control flow constructs by launching a separate dataflow job after every control flow decision (e.g., for every step of a loop). The performance of this approach is suboptimal, because (a) launching a dataflow job incurs scheduling overhead; and (b) it prevents certain optimizations across iteration steps.

In this paper, we introduce Labyrinth, a method to compile programs written using imperative control flow constructs to a single dataflow job, which executes the whole program, including all iteration steps. This way, we achieve both efficiency and ease of use.
We also conduct an experimental evaluation, which shows that Labyrinth has orders of magnitude smaller per-iteration-step overhead than launching new dataflow jobs, and also allows for significant optimizations across iteration steps.

\end{abstract}

\settopmatter{printfolios=true} %

\maketitle


\section{Introduction}

Most advanced data analysis algorithms in areas such as graph analytics and machine learning iteratively refine a dataset until a fixpoint or other convergence criteria are met. Furthermore, complex data processing pipelines, such as end-to-end machine learning, often require more general control flow, such as nested loops or if-statements. %

Parallel dataflow systems \cite{isard2007dryad, chambers2010flumejava, zaharia2012resilient, alexandrov2014stratosphere} provide a computation model that is based on executing dataflow jobs. Dataflow jobs are represented as directed graphs, whose edges represent data ``flowing'' between nodes, and whose nodes represent the computations that the system performs on this data. Dataflow systems execute these jobs in a scalable way by distributing many physical instances of each logical dataflow node across a cluster of machines. In a typical architecture, the system is running on a cluster of machines, and a client (or driver) program submits dataflow jobs\footnote{While some other authors use the term ``dataflow job'' to mean one run of the client program with the sequence of all dataflow graphs submitted throughout the run, we use it synonymously with ``dataflow graph'' (similarly to the Spark documentation).} to the system.

Incorporating control flow into the parallel dataflow model has proven to be challenging. Early systems (e.g., DryadLINQ \cite{yu2008dryadlinq}, FlumeJava \cite{chambers2010flumejava}, Apache Spark \cite{zaharia2012resilient}) do not natively support control flow inside their dataflows. Instead, control flow is expressed in the client program. Each time the client program makes a control flow decision, it submits a dataflow job to the system\footnote{The astute reader might note that even in these systems, lazy evaluation can sometimes allow for building a larger dataflow job across several control flow decisions (e.g., ``unrolling'' a loop). However, in these systems this is possible only if the control flow decisions do not depend on data that is computed by the dataflows.}. For example, each step of an iteration is typically a separate dataflow job. %

Newer systems (e.g., Apache Flink \cite{ewen2012spinning}, Naiad \cite{murray2013naiad}, TensorFlow \cite{yu2018dynamic}) employ a different approach: they allow for expressing control flow \emph{inside} their (cyclic) dataflow jobs. This has significant performance advantages over launching separate dataflow jobs: first, it eliminates the scheduling overhead from iteration steps, resulting in substantial performance gains for programs with many iteration steps; second, dataflow jobs spanning the entire program execution allow the operators of the dataflows to keep internal state between iteration steps, allowing for optimizations such as loop-invariant hoisting inside operators; third, the execution of different iteration steps can sometimes overlap, allowing for better CPU utilization and pipelined data transfer between iteration steps.

However, a current drawback of this approach is that the user has to express control flow in a functional-style API: each control flow construct is a higher-order function, to which the user has to give such functions that build the dataflows corresponding to loop bodies, loop exit conditions, branches of if-statements, etc.
We believe that for users experienced with languages such as Python, R, or Matlab, the more natural way to specify control flow is through imperative control flow constructs: while-loops, if-statements, and mutable variables that change their values during the program execution.

\begin{table}[]
  \centering
  {
  \renewcommand{\arraystretch}{1.9}
  \begin{tabular}{ l l | c | c }
    & \multicolumn{3}{c}{\hspace{7.75em}$\xlongrightarrow{\text{\normalsize{Performance}}}$} \\
    & & Out-of-dataflow & In-dataflow \\
    \cline{2-4} %
    \parbox[t]{2em}{\multirow{3}{*}{\rotatebox[origin=c]{90}{\hspace{3em}%
    $\xlongleftarrow{\text{\normalsize{Ease of use}}}$%
    }}} & Functional &  & \makecell{\vspace{-9pt}\\Flink, Naiad,\\TensorFlow} \\
    \cline{2-4} %
    & Imperative & \makecell{\vspace{-9pt}\\DryadLINQ,\\FlumeJava, Spark,\\SystemML \cite{boehm2016systemml}} & Labyrinth \\
  \end{tabular}
  \vspace{1em}
  \caption{Control flow handling approaches in parallel dataflow systems. In-dataflow control flow execution allows for better performance than executing control flow out-of-dataflow (in the client program). Imperative control flow APIs are easier to use than the functional ones. Labyrinth incorporates both of these advantages.
  }
  \label{tab:approaches}
  }
\end{table}

In this paper we are proposing Labyrinth---a system that allows the user to express control flow by easy-to-use imperative control flow constructs, while still executing it efficiently by compiling the program into a single dataflow job. For an overview of the above control flow handling approaches, see \autoref{tab:approaches}.

To be able to easily handle a wide range of imperative control flow constructs, we exploit static single assignment form (SSA) \cite{rastello2016ssa}, which is a popular intermediate representation of control flow in compilers of imperative languages. SSA is able to represent a wide range of control flow constructs: loops, if-statements, and even unstructured control flow, such as break, continue, or goto. By utilizing SSA, we avoid devising separate compilation procedures for each control flow construct, and instead handle them in a unified way in our compilation and coordination.

We also present an algorithm for coordinating the distributed execution of control flow. Even though TensorFlow and Naiad also have mechanisms for this purpose, our algorithm is tailored to SSA. This allows us to rely on the concepts of SSA end-to-end, and thereby makes the compilation from SSA to dataflows more straightforward. Our compilation creates dataflows whose structures mirror the structure of SSA: we create dataflow nodes from variable assignments, and dataflow edges from variable references (even across basic block boundaries).

In sum, we make the following contributions:

\begin{enumerate}

\item To provide easy-to-use imperative control flow constructs and still execute programs efficiently, we show how to build a single dataflow job from a program in SSA representation. %

\item We propose an algorithm for efficiently coordinating the distributed execution of control flow in our dataflows. The algorithm is tailored to the SSA representation of control flow. %

\item We incorporate two exemplary optimizations into our dataflows: a)~exploiting loop-invariant datasets for reusing the build-side of a hash join across all steps of an iteration, and b)~loop pipelining. These optimizations are made possible by executing the entire program in a single dataflow job.

\item We experimentally verify that compiling a program that has control flow to a single dataflow job results in substantial performance gains over launching separate dataflow jobs after every control flow decision.

\end{enumerate}

The rest of the paper is structured as follows: \autoref{sec:background} presents important concepts we build upon, and \autoref{sec:overview} provides an overview of the current approaches for handling control flow in dataflow systems.
\autoref{sec:solution-overview} gives an overview of our system.
\autoref{sec:ssa-to-single-dataflow} discusses how we can compile any program from SSA representation into Labyrinth dataflows.
\autoref{sec:distr-coord} presents a mechanism for the efficient coordination of the distributed execution of control flow.
\autoref{sec:loop-invariant-runtime} shows how we can apply loop-invariant hoisting for hash joins.
\autoref{sec:impl} discusses some implementation details.
\autoref{sec:eval} presents our evaluation, \autoref{sec:related_work} compares our approach with related work, and \autoref{sec:conclusion} concludes.

\section{Background Concepts} \label{sec:background}

In this section, we first introduce basic concepts used throughout this paper, including SSA. Then, we briefly describe the collection-based APIs of modern dataflow systems. Note that we use the phrase \emph{iteration step} to denote one execution of a loop body, and we use \emph{iteration} as a synonym of \emph{loop}.

\subsection{Control Flow Analysis} %

A \emph{basic block} \cite{aho2007compilers} is a maximal continuous sequence of instructions which always execute one after the other. This means that the execution can only enter a basic block at its beginning and only leave it at its end. For example, the body of a while-loop is a basic block if there are no other control flow instructions inside. %

The \emph{control flow graph} \cite{aho2007compilers} of a program is a graph whose nodes correspond to the basic blocks of the program, and whose edges show the possible control flow paths. Specifically, let $u$ and $v$ be two nodes, with corresponding basic blocks $U$ and $V$. A directed edge goes from $u$ to $v$, if control flow can directly go from $U$ to $V$. For example, in case of an if-statement, edges go from the node of the basic block before the if-statement to the nodes of the basic blocks of the then- and else-branches.

A \emph{control flow merge point} \cite{aho2007compilers} is a basic block whose corresponding node in the control flow graph has an in-degree of at least two. For example, the basic block after the two branches of an if-statement is a merge point, because at this point the different possible paths merge.

\subsection{Static Single Assignment Form} \label{sec:SSA}

SSA is a widely used intermediate representation in compilers \cite{rastello2016ssa}. The basic defining characteristic of SSA is that there is a one-to-one correspondence between variables and assignment statements. An important consequence of this is \emph{referential transparency}: every variable reference refers to a value that was written to the variable at its unique assignment statement. In contrast, when a program is not in SSA, the value of a variable reference depends on its context:
\begin{enumerate}
\item It can depend on the position of the variable reference. Specifically, which of the assignment statements to the variable precede the reference in the program text.
\item Which of the preceding assignments was executed at run-time may depend on how the control flow proceeded.
\end{enumerate}

When there is no control flow in the program, only issue (1) can arise. In this case, we can transform to SSA, and thus eliminate the problem by
(a) changing the variable names on the left-hand sides of assignment statements so that each of them assigns to unique variables, and
(b) changing the variable references appropriately, so that they reference the variable of the most recent assignment.
We can see an example of this in \autoref{fig:ssa-example1}, where the variable $a$ is split into $a_1$ and $a_2$.

\algrenewcommand\algorithmicindent{0.8em}

\begin{lstfloat}[t]
\small
  \begin{sublstfloat}{95pt}
    \begin{tabular}{ l | r }
      \begin{minipage}{29pt}
        $a = 0$\\
        $b = a + 1$\\
        $a = 5$\\
        $c = a + 1$
      \end{minipage}
      &
      \begin{minipage}{38pt}
        $a_1 = 0$\\
        $b_1 = a_1 + 1$\\
        $a_2 = 5$\\
        $c_1 = a_2 + 1$
      \end{minipage}
    \end{tabular}
      \caption{ }
        \label{fig:ssa-example1}
  \end{sublstfloat}
  \begin{sublstfloat}{118pt}
    \begin{tabular}{ l | r }
    \begin{minipage}{53pt}
      \begin{algorithmic}
        \State $a = 0$
        \If{...}
          \State $a = a + 1$
          \Else
          \State $a = a + 2$
        \EndIf
        \State $b = a + 5$
      \end{algorithmic}
    \end{minipage}
    &
    \begin{minipage}{65pt}
      \begin{algorithmic}
        \State $a_1 = 0$
        \If{...}
          \State $a_2 = a_1 + 1$
        \Else
          \State $a_3 = a_1 + 2$
        \EndIf
        \State $a_4 = \Phi(a_2, a_3)$
        \State $b_1 = a_4 + 5$
      \end{algorithmic}
    \end{minipage}
    \end{tabular}
    \caption{ }
      \label{fig:ssa-example2}
  \end{sublstfloat}
  \caption{Two programs with their SSA representations.}
  \label{fig:ssa-examples}
\end{lstfloat}

However, if there is control flow in the program, then issue (2) can also occur. In this case, after performing the change of variable names on the left-hand sides of assignments, we cannot change the references in the above-described way. This is because variable references at control flow merge points can refer to the value assigned in either one of the possible control flow paths. SSA resolves this problem by introducing a new variable at a merge point, and using a so-called $\Phi$-function to assign a value to this new variable. The only purpose of these $\Phi$-functions is to disambiguate these references, i.e., to choose one of their inputs based on how the control flow actually proceeded. We can see an example in \autoref{fig:ssa-example2}.

\subsection{Collection-Based Data Processing APIs} \label{sec:assumptions-src-lang}

Modern dataflow systems provide collection-based APIs on top of the dataflow-graph-based APIs \cite{yu2008dryadlinq, chambers2010flumejava, zaharia2010spark, alexandrov2014stratosphere}. The programmer can use a parallel collection type for those collections that are too large to be processed on a single machine, and therefore must be handled in a scalable way. %
This relieves the user from thinking in terms of the dataflow graph, which is instead built by the system from the operations that were specified on the parallel collections (e.g., RDD in Spark).

In our system, the language that we are compiling from has a type named \emph{Bag}, which is our parallel collection abstraction. It is a multiset, i.e., an unordered collection of elements, where duplicates are allowed. We provide a set of operations on bags that are commonly found in parallel dataflow systems, such as Flink or Spark.

\section{Handling Control Flow in Current Dataflow Systems} \label{sec:overview}

In this section, we demonstrate through an example what kind of data analysis programs benefit from imperative control flow constructs. Then, we discuss current systems' control flow handling and their shortcomings. We also compare the functional control flow API approach with the imperative approach.

\subsection{Example Program} \label{sec:example}

\algrenewcommand\algorithmicindent{0.61em}

\definecolor{pageAttrColor}{HTML}{009900}
\definecolor{commentColor}{HTML}{707070}

\begin{lstfloat*}[t]

  \begin{sublstfloat}{0.57\textwidth}
    {\small
      \begin{algorithmic}[1]
      \State {\color{pageAttrColor} pageAttributes} = readFile(``pageAttributes'')
      \State initialDay = SingletonBag(1) {\color{commentColor} // ``1'' lifted to be a bag}
      \State initialCounts = EmptyBag %
      \State whileLoop( {\color{commentColor} // Higher-order function call}
        \State \hspace{0.5em} {\color{commentColor} // First two arguments are the initial values of the loop variables:}
        \State \hspace{0.5em} initialDay, initialCounts
        \State \hspace{0.5em} {\color{commentColor} // Third arg is the function building the dataflow for the exit cond:}
        \State \hspace{0.5em} (day, yesterdayCounts) => \{
          \State \hspace{1em}   day $\leq$ SingletonBag(365) {\color{commentColor} // (``$\leq$'' is a bag operation here)}
        \State \hspace{0.5em} \}
        \State \hspace{0.5em} {\color{commentColor} // Fourth arg is the function building the dataflow for the body:}
        \State \hspace{0.5em} (day, yesterdayCounts) => \{
          \State \hspace{1em} fileName = day.map(d => ``pageVisitLog'' + d)
          \State \hspace{1em} visits = readFile(fileName)
          \State \hspace{1em} visits = visits.join({\color{pageAttrColor} pageAttributes}).filter(p $=>$ p.type = ...)
          \State \hspace{1em} counts = visits.map(x $=>$ (x,1)).reduceByKey(\_ + \_)
          \State \hspace{1em} if( {\color{commentColor} // Higher-order function call}
            \State \hspace{1.5em} {\color{commentColor} // First arg is the function building the dataflow for the condition:}
            \State \hspace{1.5em} () => day != SingletonBag(1), {\color{commentColor} // ``1'' lifted to be a bag}
            \State \hspace{1.5em} {\color{commentColor} // 2nd arg is the function building the dataflow for the then-branch:}
            \State \hspace{1.5em} () => (counts join yesterdayCounts)
              \State \hspace{2.5em} .map((id,today,yesterday) $=>$ abs(today - yesterday))
              \State \hspace{2.5em} .reduce(\_ + \_).writeFile(``diff'' + day)
          \State \hspace{1.5em})
          \State \hspace{1em} day = day.map(d => d + 1)
          \State \hspace{1em} {\color{commentColor} // The loop body function returns the next values of the loop vars:}
          \State \hspace{1em} (day, counts)
        \State \hspace{0.5em} \}
        \State )
      \end{algorithmic}
      \caption{Functional control flow. This imagined API is an ``idealized'' hybrid between the APIs of Naiad and TensorFlow, in the sense that we have combined their advantages and also removed some accidental complexity, see \autoref{sec:func-api}.}
      \label{fig:example-src-func}
    }
  \end{sublstfloat}
  \begin{sublstfloat}{0.42\textwidth}
       {\small
       \begin{algorithmic}[1]
              \State {\color{pageAttrColor} pageAttributes} = readFile(``pageAttributes'')
              \State yesterdayCounts = null %
              \State day = 1
              \While{day $\leq$ 365}
                     \State {\color{commentColor} // Read all page-visits for this day}
                     \State visits = readFile(``pageVisitLog'' + day) {\color{commentColor} // integer pageIDs}
                     \State {\color{commentColor} // We want to examine only pages of a certain type, so}
                     \State {\color{commentColor} // we get the page types from a large lookup table:}
                     \State visits = visits.join({\color{pageAttrColor} pageAttributes}).filter(p $=>$ p.type = ...)
                     \State {\color{commentColor} // Count how many times each page was visited:}
                     \State counts = visits.map(x $=>$ (x,1)).reduceByKey(\_ + \_)
                     \State {\color{commentColor} // Compare to previous day (but skip the first day)}
                     \If{day != 1}
                            \State diffs =\\
                              \hspace{2.5em} (counts join yesterdayCounts)\\
                              \hspace{2.5em} .map((id,today,yesterday) $=>$ abs(today - yesterday))
                            \State diffs.reduce(\_ + \_).writeFile(``diff'' + day)
                     \EndIf
                     \State yesterdayCounts = counts
                     \State day = day + 1
              \EndWhile
       \end{algorithmic}
       }
       \caption{Imperative control flow (Labyrinth).}
       \label{fig:example-src-imp}
  \end{sublstfloat}

  \caption{A comparison of control flow APIs through an example data analysis program, which compares statistics of consecutive days from a year of page visit logs. \emph{PageAttributes} (marked with {\color{pageAttrColor} green}) is loop-invariant.} %

  \label{fig:example-src-comparison}

\end{lstfloat*}

In \autoref{fig:example-src-imp}, we can see an example data analysis program expressed using imperative control flow constructs, i.e., a while-loop, if-statement, and mutable variables. 
The goal of the analysis is to identify days of anomalous traffic during the last year in a page visit log. We keep the logs in separate files for each day and we compare some statistics of consecutive days. Each log entry is an integer page ID, meaning that someone visited the specified page. For each day, we compute a collection of separate visit counts for each page, i.e., a collection of (\emph{PageID}, \emph{visit-count}) pairs. %

The program operates as follows. A loop iterates through the days of the year. Each iteration step starts with reading the log file of the day (Line 6). Let us imagine that we are interested only in anomalies about a certain type of the pages, but the logs do not contain information about the page type. Therefore, we perform a join with the dataset of page types, and then filter based on page type (Line 9). Then, we compute the visit counts of the pages with a \emph{reduceByKey} operation (Line 11) (akin to the well-known word count example). Then, if the day of the current iteration step is not the first day, we compare the visit counts of today with yesterday, and output the sum of the differences (Lines 13--18).

\subsection{Separate Dataflow Jobs for Each Step}

\begin{figure}
  \centering
    \includegraphics[width=0.42\textwidth]{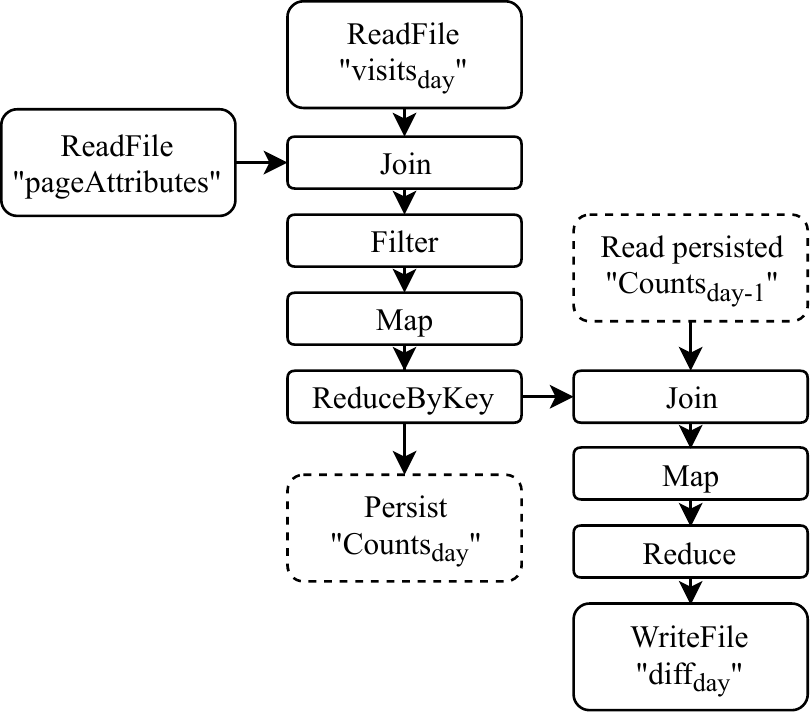}
  \caption{Dataflow for one iteration step of the example program in \autoref{fig:example-src-comparison}.}
  \label{fig:example-dataflow-one-step}
\end{figure}

In Spark and Flink, the above example program cannot be executed as a single dataflow job\footnote{Flink allows for control flow inside dataflows only in the case of fixpoint iterations.}. Instead, the user has to launch a separate job for every step of the iteration.
We can see such a job in \autoref{fig:example-dataflow-one-step}. As we are comparing visit counts from different days, we have to always persist the computed counts (at least to memory), and then read this persisted dataset again in the job of the next day. Note that the burden of knowing that the dataset should be persisted typically falls on the user (e.g., making a \texttt{.cache} call in Spark), as the system has no easy way of knowing whether a dataset will be needed in a later job.

\subsubsection{Scheduling Overhead} \label{sec:chall-scheduling}

Launching a dataflow job incurs the overhead of scheduling the execution of the parts of the job on the cluster. As we show in \autoref{sec:exp-sched-overhead}, this overhead can easily be as large as 250 ms even for very simple dataflows on moderately sized clusters, and increases linearly with cluster size and number of dataflow operators.

If the data involved in one iteration step is not too large, then the scheduling overhead of launching new dataflow jobs becomes the dominating component of the execution time. Note that the total amount of data in all iteration steps can still be large, in which case a non-parallel execution can still take a long time. For example, the program in \autoref{fig:example-src-comparison} has 365 iteration steps. %

Labyrinth incorporates all iteration steps into a single cyclic dataflow job. This allows for reusing the same physical operator instances across iteration steps, thereby eliminating the scheduling overhead.

\subsubsection{Missed Optimization Opportunity: Loop-Invariant Hoisting} \label{sec:static}

Iterations often involve some loop-invariant (static) datasets, which are reused without updates during subsequent iteration steps. We can see an example of this in \autoref{fig:example-src-comparison}, where the \emph{pageAttributes} dataset (marked green) is read from a file outside the iteration and is used in a join inside the iteration in Line 8. %

It is a common optimization to pull those parts of a loop body that depend on only static datasets outside of the loop, and thus execute these parts only once \cite{ewen2012spinning, bu2010haloop, zhang2012imapreduce, ekanayake2010twister}.
However, launching new dataflow jobs for every iteration step prevents this optimization in the case of such binary operators where only one input is static.
For example, if a static dataset is used as the build-side of a hash join, then the system should not rebuild the hash table at every iteration step. Labyrinth operators can keep such a hash table in their internal states between iteration steps. This is made possible by implementing iterations as a single cyclic dataflow job, where the lifetimes of operators span all the steps.

\subsection{Functional Control Flow Constructs} \label{sec:func-api}

Flink, Naiad, and TensorFlow \cite{yu2018dynamic} support control flow inside dataflow jobs. However, they rely on functional-style APIs, where each control flow construct is a higher-order function. For example, \texttt{tf.while\_loop} is a higher-order function, which expects (among other things) two functions: one for building the dataflow of the loop body and an other for building the dataflow of the loop exit condition\footnote{A search for the term \texttt{tf.while\_loop} on stackoverflow.com or the TensorFlow mailing lists shows that many users are confused by this API.}.

We can see our example program expressed using an imagined functional control flow API in \autoref{fig:example-src-func}. In the imagined API, we combined the advantages of Naiad's and TensorFlow's APIs, and tried to simplify as much as possible by removing any accidental complexity.
For example, the imagined API uses Scala-like pseudocode, but TensorFlow uses Python, which does not allow multi-line lambdas, and therefore the loop body and other functions would have to be written as separate functions and referenced in the higher-order function calls, making the program structure harder to parse by humans.
The API is assumed to have joins, which TensorFlow does not have, as it focuses on machine learning.
Naiad's API has the practical issue that its loop construct allows for only one loop variable (more can be added by tedious low-level graph-building), but we imagine that this limitation could be removed. %

In these functional APIs, somewhat counter-intuitively, a user function for a loop body is always executed \emph{exactly once}, as it only builds a dataflow instead of doing the actual work. This invites mistakes: when writing such a loop body function, users have to pay attention to not accidentally write code that would directly (in the client program) do work that is intended for every iteration step. Instead, code in loop body functions should build a representation of the work into the dataflow job. %

Note that in the dataflow-building functions that we give to the control flow functions, we have to make even the non-collection values part of the dataflow job. (To see why, keep in mind that these functions are always executed exactly once, even for loop bodies.) For example, even though the \emph{day} variable holds just one number (the number of the current day), we have to ``lift'' it into the dataflow job by creating a single-element bag for it. (These are expressed as rank 0 tensors in TensorFlow.) Labyrinth performs this lifting automatically, see \autoref{sec:uniform}.

\section{Labyrinth Overview} \label{sec:solution-overview}

In this section, we provide an overview of our compiler pipeline. It compiles programs from a high-level data analytics language having a collection-based API and imperative control flow constructs into cyclic dataflows of Labyrinth. Additionally, we illustrate the control flow handling in Labyrinth by an example.

\subsection{The Compiler Pipeline} \label{sec:compiler-pipeline}

We provide an overview of the architecture of the system in \autoref{fig:architecture}. Labyrinth compiles from imperative control flow to dataflows in two steps:

\begin{figure}
  \centering
    \includegraphics[width=0.48\textwidth]{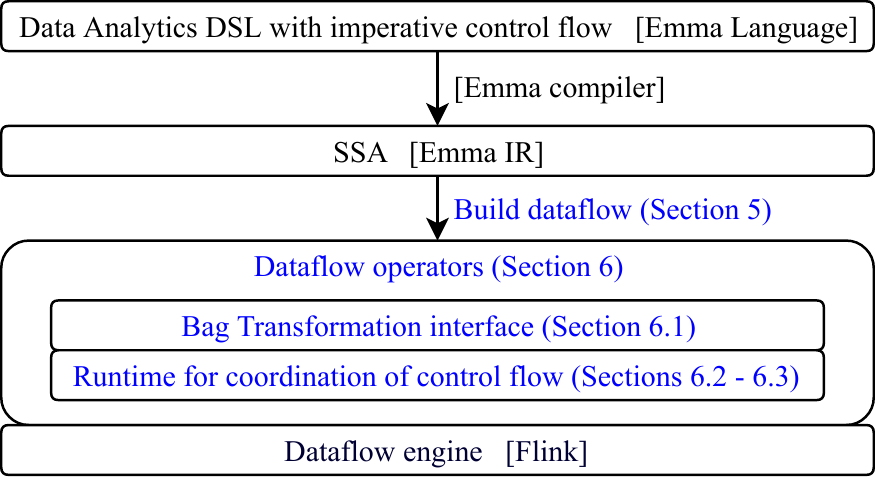} %
  \caption{Overview of the system architecture. Components that are the focus of this paper are {\color{blue} blue}.
  The concrete implementations of components are shown in [brackets], and are briefly discussed in \autoref{sec:impl}.
  }
  \label{fig:architecture}
\end{figure}

\textbf{Compiling to SSA.}
The first step of the compilation procedure is to transform the imperative program into a typed SSA representation. There are well-known algorithms for building an SSA representation \cite{rastello2016ssa}. Thus, in this paper we focus on the rest of the compilation pipeline and on the coordination of the distributed execution of control flow.%

\textbf{Compiling SSA to Dataflow.}
We use SSA as our intermediate representation, because it makes data dependencies explicit in a way that fits naturally to the dataflow model. On the one hand, the (unique) assignment statement of each variable can be directly translated into a dataflow node. Moreover, variable references on the right-hand side of an assignment show what input data is needed by the corresponding node, and therefore can be directly translated into dataflow edges.

\subsection{An Example Dataflow in Labyrinth}
\definecolor{ifCondColor}{HTML}{0000CC}
\definecolor{exitCondColor}{HTML}{991100} %
\algrenewcommand\algorithmicindent{0.9em}
\begin{figure}[] %
  \begin{subfigure}{0.5\textwidth}
       {%
       \begin{algorithmic}[1]
              \State \tikzmark{start1}pageAttributes = readFile(``pageAttributes'')
              \State yesterdayCnts$_1$ = null %
              \State day$_1$ = 1\tikzmark{end1}
              \Do
                     \State \tikzmark{start2}yesterdayCnts$_2$ = $\Phi$(yesterdayCnts$_1$,\color{exitCondColor}yesterdayCnts$_3$\color{black})
                     \State day$_2$ = $\Phi$(day$_1$,\color{exitCondColor}day$_3$\color{black})
                     \State fileName = ``pageVisitLog'' + day$_2$
                     \State visits$_1$ = readFile(fileName)
                     \State joinedWithAttrs = visits$_1$.join(pageAttributes)
                     \State visits$_2$ = joinedWithAttrs.filter(p $=>$ p.type = ...)
                     \State visitsMapped = visits$_2$.map(x $=>$ (x,1))
                     \State counts = visitsMapped.reduceByKey(\_ + \_)
                     \color{ifCondColor}
                     \State ifCond = day$_2$ != 1\tikzmark{end2}  %
                     \color{black}
                     \If{ifCond}
                            \State \tikzmark{start3}joinedYesterday = \color{ifCondColor} counts \color{black} join \color{ifCondColor} yesterdayCnts$_2$ \color{black}
                            \State diffs = joinedYesterday.map(...)
                            \State summed = diffs.reduce(\_ + \_)
                            \State outFileName = ``diff'' + day$_2$
                            \State summed.writeFile(outFileName)\tikzmark{end3}
                     \EndIf
                     \State \tikzmark{start4}yesterdayCnts$_3$ = counts
                     \State day$_3$ = day$_2$ + 1
                     \color{exitCondColor}
                     \State exitCond = day$_3 \le 365$\tikzmark{end4} %
                     \color{black}
              \DoWhile{exitCond}
       \end{algorithmic}
       } 
       \centering \caption{}
  \end{subfigure}
  
  \vspace{1em}

  \begin{subfigure}{0.47\textwidth}
    \vspace{1em}
    \includegraphics[width=\textwidth]{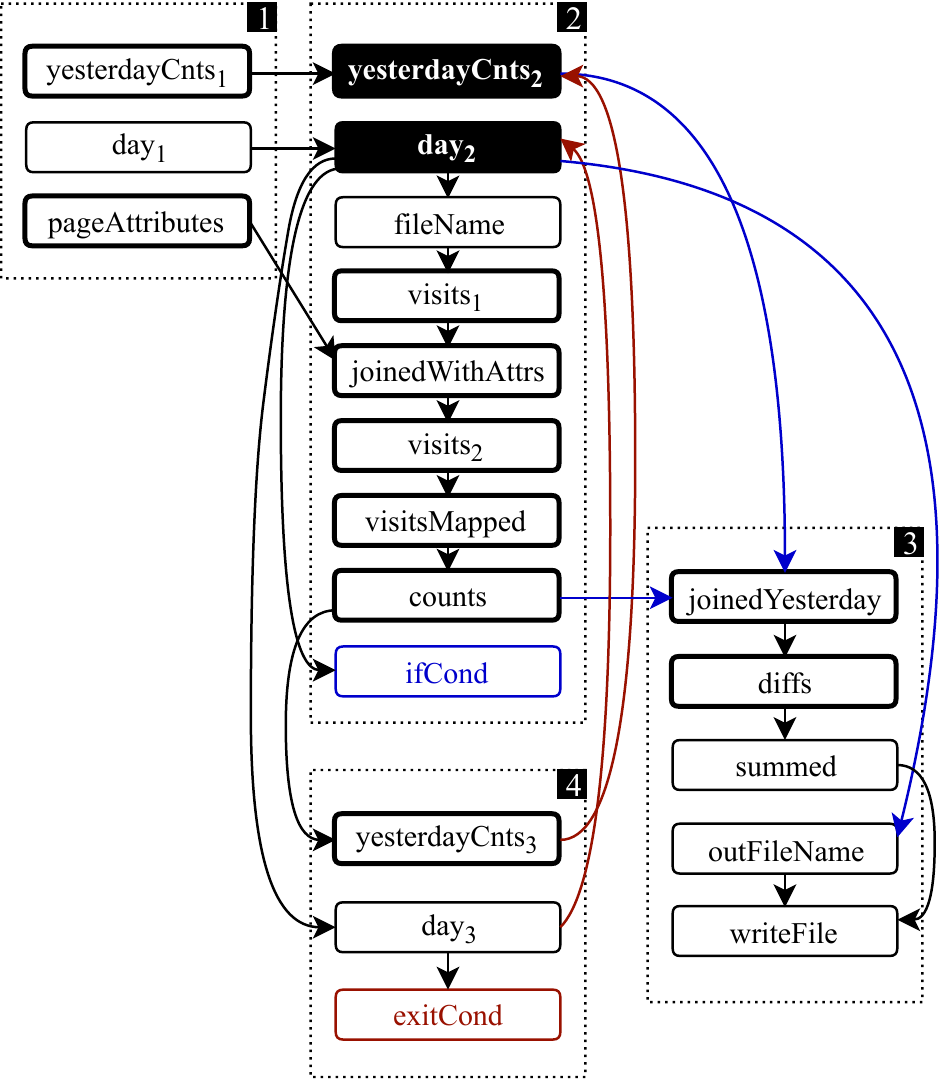}
    \centering \caption{}
  \end{subfigure}

  \caption{The program in \autoref{fig:example-src-imp} transformed to SSA representation (a), and its Labyrinth dataflow (b). The basic blocks are marked with dotted rectangles. The smaller, rounded rectangles are dataflow nodes, corresponding to variables in the SSA representation.
  The variables corresponding to the thick bordered nodes have bag types.
  The colors are explained in \autoref{sec:topology}.
  }
  \label{fig:example-ssa-and-dataflow}

  \BasicBlock[19em][0.2em]{start1}{end1}{1}%
  \BasicBlock[14em][0.2em]{start2}{end2}{2}%
  \BasicBlock[7.55em][0.2em]{start3}{end3}{3}%
  \BasicBlock[12.45em][0.2em]{start4}{end4}{4}%

\end{figure}

\algrenewcommand\algorithmicindent{1.4em}%
In this section, we provide a brief overview of how Labyrinth works, using the example program in \autoref{fig:example-src-imp}.
We can see the program's SSA representation and Labyrinth dataflow in \autoref{fig:example-ssa-and-dataflow}.
Many of the nodes in the Labyrinth dataflow correspond to similar nodes in \autoref{fig:example-dataflow-one-step}, which shows a dataflow for one step of the iteration. However, Labyrinth adds several types of extra nodes, which we briefly discuss in the following paragraphs.

\textbf{Nodes for Non-Bag Variables.} In such dataflow systems where control flow is executed in the client program, variables that are not bags (e.g., the loop counter) are typically not included in dataflows. However, Labyrinth represents the entire program in a single dataflow job, and therefore has to create dataflow nodes also from these variables. These nodes are the non-thick bordered ones in \autoref{fig:example-ssa-and-dataflow}.

\textbf{Condition Nodes.} We create a node for the \emph{ifCond} variable (Line 13), similarly to the other variables. However, this node has an additional task: it broadcasts the result of evaluating the if-condition to all other nodes. From this, the nodes whose variables are inside the then-block of the if-statement know whether they should expect any input data, and their input nodes know whether to send data to them. We also add a condition node for the exit condition of the loop (Line 23), which informs the other nodes whether to end the iteration or start a new step.

\textbf{$\Phi$-Nodes.} The nodes with inverted colors are $\Phi$-nodes. Unlike other nodes, the origins of their inputs depend on the execution path the program took so far: in the first iteration step, they get their values from outside the iteration (Lines 2--3), but later from the previous step (Lines 21--22). This choice is represented by the $\Phi$-functions of the SSA form.

\section{Compiling SSA to Dataflow} \label{sec:ssa-to-single-dataflow}

In this section, we describe how we compile from SSA to Labyrinth dataflows. %

\subsection{Assumptions} \label{sec:compilation-assumptions}

\textbf{Source Language.}
Our approach assumes that the data analytics language that we are compiling from has operations that can be translated to parallel dataflows, and it has imperative control flow constructs that can be translated to SSA. 
The control flow structure of programs should be visible to the system.
This can be achieved by an embedded domain-specific language (DSL) that has a deep enough embedding that exposes control flow instructions, such as the Emma language \cite{alexandrov2015implicit, alex2018representations} or a DSL that is using the Lightweight Modular Staging approach \cite{rompf2012lightweight}. An external DSL, such as the language of SystemML \cite{ghoting2011systemml, boehm2016systemml}, could also be the source language since there the entire source code is visible to the system. %
We also assume that the order of any side effects, such as file IO, do not matter.\footnote{Alternatively, we could enforce the ordering of IO operations by adding extra dataflow edges between them (similarly to \texttt{tf.control\_dependencies}).}

\textbf{Dataflow Engine.}
We require the following commonly occurring features
from the dataflow engine that we are targeting in our compilation. Client programs can submit dataflow jobs to a cluster of machines, where worker processes of the dataflow engine are running. These jobs are specified through a \emph{logical dataflow graph} with arbitrary stateful computations in the vertices and pipelined data transfer on the edges. The dataflow engine translates the logical dataflow graph into a \emph{physical dataflow graph}. This involves creating multiple \emph {physical instances} of each logical node and distributing them to the worker machines in the cluster. Note that many systems only allow for acyclic dataflow graphs, but we also require support for adding arbitrary cycles to the graph. Examples of systems that support this requirement are Flink \cite{alexandrov2014stratosphere}, Naiad \cite{murray2013naiad}, and TensorFlow \cite{abadi2016tensorflow}.

\textbf{Intermediate Representation.}
We show the part of the compilation procedure that starts from an SSA representation. We also assume that every intermediate value that occurs in the program is assigned to a variable. This means that the right-hand side of every assignment is either a constant or a function call whose arguments cannot be complex expressions, but can only be references to variables. For example, rewriting the assignment statement $a = f(g(b),h(c))$ to conform to this rule results in the following code:
$ x = g(b); $ \hspace{0.1em} $ y = h(c); $ \hspace{0.1em} $ a = f(x,y) $.

\subsection{Lifting non-Bag Variables} \label{sec:uniform}

The variables that are our primary concern are those that hold bags, since these contain most of the data (see \autoref{sec:assumptions-src-lang}). Nevertheless, since we would like to compile the entire program into a single dataflow job, our compilation process has to include all other variables as well, i.e., variables with non-bag types, such as loop counters. (These have a non-thick border in \autoref{fig:example-ssa-and-dataflow}.)

We handle bags and other variables in a uniform way in the following sections. To achieve this, as a first compilation step, we wrap non-bag values in one-element bags and appropriately change the operations on them:

\begin{itemize}%
\item if a variable has a non-bag type \texttt{A}, we change it to \texttt{Bag<A>};
\item an I/O operation of a non-bag value is changed to an equivalent bag operation;
\item a unary function $f$ that acts on a non-bag value is wrapped in a \emph{map} transformation, whose user-defined function (UDF) is $f$;
\item a binary function
$f$ that acts on two non-bag values is substituted by a \emph{cross} and a \emph{map}. The \emph{cross} creates a one-element bag that contains a pair whose two components are the elements of the two input bags. The \emph{map} operates on this pair and has $f$ as its UDF.
\end{itemize}
After the above changes, all variables in the program are bags.

Note that we assume that most of the data is in the variables that originally had a bag type, and therefore non-bag variables do not significantly affect performance.

\subsection{Building the Dataflow Graph} \label{sec:topology}

As mentioned before, our dataflows mirror the SSA representation, by translating each variable to a dataflow node and each variable reference to a dataflow edge. More formally, for each variable $a$, we create a dataflow node $n(a)$. The assignment statement\footnote{Recall that in SSA each variable has exactly one assignment statement to it.} to $a$ determines the input edges of $n(a)$: for each variable~$b$ on the right-hand side, we create an edge from $n(b)$ to $n(a)$.

We will call certain edges \emph{conditional output edges}. These have the property that not all output bags of their source nodes should be sent to their target nodes. This can happen when the nodes are in different basic blocks, or when there is a $\Phi$-function at the beginning of a loop body receiving input from a variable defined later in the loop body.
In \autoref{fig:example-ssa-and-dataflow} conditional output edges are blue or red.

If a node has one or more conditional output edges, then it needs some additional information to determine for a particular output bag, on which conditional output edges (zero, one, or multiple) to send that bag.
Because every intermediate value is assigned to a variable (see \autoref{sec:compilation-assumptions}), every boolean condition that determines the actual control flow (i.e., if-conditions and loop-conditions) is only a reference to a boolean variable (and not a complex expression). The nodes of these variables compute whether the conditions hold, and we call these nodes \emph{condition nodes}. In \autoref{fig:example-ssa-and-dataflow} we show condition nodes with different colors. Conditional edges have the same color as the condition nodes that they depend on.
We show in \autoref{sec:distr-coord} how the results of evaluating the conditions are propagated to nodes that have conditional outputs depending on it.
Note that the $\Phi$-function is treated like any other bag-transformation when creating its dataflow node and connecting its edges.

The right-hand side of every assignment is a call to a primitive bag operation. Therefore, for each of our primitives, we have to give an implementation that will run inside a physical operator of the dataflow engine that we are targeting. These implementations can be similar to those found in typical dataflow systems, such as Flink or Spark: they consume partitions of input bags, and produce partitions of output bags. The difference between our operator implementations and that of other systems is that we have to handle general control flow. We show how to do this in the next section.  %
\section{Coordinating the Distributed Execution of Control Flow}  \label{sec:distr-coord}

\gabornote{I also realized that I used the word ``operator'' ambiguously: sometimes I meant the thing that implements the interface in \autoref{sec:operator-interface}, and sometimes I meant the thing that wraps the previous thing, i.e., a low-level Flink operator that does the job of the runtime. I guess clearing this up also increases the readability of this section. I changed the latter to ``transformation'' (I didn't mark these changes red)}

In this section, we describe Labyrinth's mechanism for coordinating the distributed execution of control flow. We achieve a separation of concerns by first defining an interface for bag-transformations. Implementations of this interface do not need to deal with control flow, but can concentrate on the semantics of the particular bag-transformation. Then, in the rest of the section, we show how to deal with control flow. That is, we bridge the gap between implementations of the bag-transformation interface operating on the level of bags, and an existing parallel dataflow execution engine, such as Flink.

\subsection{Bag-Transformation Interface} \label{sec:operator-interface}

Conceptually, the semantics of a bag-transformation are specified by a function that has one or two inputs of type bag and outputs one bag. However, we aim for pipelined data transfer between transformations, which means that our transformation implementations have to work on one element at a time %
instead of whole input bags.

Our transformation interface is very similar to other dataflow frameworks that provide a push-based interface.
The interface has the following methods to be implemented by every transformation (binary transformations have two \texttt{PushInElement} and \texttt{CloseInBag} methods):
\begin{itemize}%

\item \texttt{OpenOutBag}: The runtime calls this method on each physical instance of the transformation when the computation of a new output bag should be started. The transformation can initialize its state here.

\item \texttt{PushInElement}: The runtime calls this method to give an element of the current input bag to the transformation.

\item \texttt{CloseInBag}: %
The runtime calls this method when no more elements of the current input bag will arrive at this physical instance of the transformation.
For example, a grouped aggregation can emit the final aggregates from its hash table here. %

\end{itemize}

The runtime also provides an output collector object to transformations for emitting output. This object has two methods: \texttt{Emit}, for emitting an element of the output bag to downstream transformations, and \texttt{CloseOutBag}, for telling the runtime that the current output bag partition is closed, i.e., no more elements of the current bag will be produced by this physical instance of the transformation.

The runtime takes care of coordinating control flow and separating the computations of the multiple output bags of a transformation that is reached multiple times throughout the program execution. In particular, the elements provided through \emph{PushInElement} all belong to those input bags that should take part in the computation of the current output bag. Thus, a transformation has to deal with computing just one output bag at a time. %

From now on, we use the word ``transformation'' to refer to implementations of this interface, and we use the word ``operator'' to refer to an operator at the level of the execution engine (e.g., Flink). Our operators are composed of a transformation plus the control flow handling support of the Labyrinth runtime.

\subsection{Challenges for the Runtime} \label{sec:chall-distr-coord}

In this section, we point out two challenges that the coordination algorithm described in the next section solves.

\algrenewcommand\algorithmicindent{1em} %

\begin{lstfloat}[t]
\small
    \begin{sublstfloat}{0.48\columnwidth} %
       \centering
       \begin{algorithmic}
              \While{...}
                     \State $x$ = ...
                     \While{...}
                            \State $y$ = ...
                            \State $z$ = $f(x,y)$
                     \EndWhile
              \EndWhile
       \end{algorithmic}
       \caption{ }
       \label{fig:nested-loops-example}
    \end{sublstfloat}
    ~
    \begin{sublstfloat}{0.48\columnwidth} %
       \centering
       \begin{algorithmic}
              \While{...}
                     \State // Basic block $A$
                     \If{...}
                            \State // Basic block $B$
                            \State $x_1$ = ...
                            \State $y_1$ = ...
                     \Else
                            \State // Basic block $C$
                            \State $x_2$ = ...
                            \State $y_2$ = ...
                     \EndIf
                     \State // Basic block $D$
                     \State $x_3$ = $\Phi(x_1,x_2)$
                     \State $y_3$ = $\Phi(y_1,y_2)$
                     \State $z$ = $f(x_3,y_3)$
              \EndWhile
       \end{algorithmic}
       \caption{ }
       \label{fig:sync-example}
    \end{sublstfloat}
    \caption{Example programs with non-trivial control flow structures.
    $f$ denotes an arbitrary binary transformation.
    }
    \label{fig:examples-control-flow}
\end{lstfloat}

\algrenewcommand\algorithmicindent{1.4em}

\textbf{1. The matching of input bags of binary transformations is not always one-to-one:}
In the case of binary transformations, the runtime gives a pair of bags to a bag-transformation (which implements the interface shown in the previous section) at a time. Therefore, we have to match bags arriving on one logical input edge to bags arriving on the other logical input edge. If both of the inputs are coming from transformations that are in the same basic block, then each bag from one input is matched to exactly one bag from the other input. However, if the transformations producing the two inputs are enclosed in different loops, as in the example program in \autoref{fig:nested-loops-example}, then the matching will not be one-to-one. Here, the runtime has to match each bag from $x$ to several bags from $y$ when feeding pairs of bags to $f$ (see also in \autoref{sec:static}). %

\textbf{2. First-come-first-served does not work for choosing the input bags to process:}
How do operators decide which input bags to process next? A simple way would be to order input bags by the arrival of their first elements. However, this does not always yield a correct execution as illustrated by \autoref{fig:sync-example}.
Suppose that the control flow reaches the basic blocks in the following order: $ABDACD$. In this case, it is possible that, due to irregular processing delays, the operator of $x_3$ gets data from $x_1$ first and then from $x_2$, while the operator of $y_3$ gets data from $y_2$ first and then from $y_1$. This can happen because the operators in the different if-branches are not synchronized, i.e., they do not agree on a global order in which to process bags. %
However, in order to get a correct result, the operator of $z$ has to match the bag that originates from $x_1$ with the bag that originates from $y_1$, and match the bag that originates from $x_2$ with the bag that originates from $y_2$. %

\subsection{Coordination Based on Bag-Identifiers} \label{sec:cfl}

The high-level structure of our solution to the above challenges is the following. We introduce a \emph{bag-identifier} (\autoref{sec:bag-ids}), which is straightforward to define and keep track of in the case when we are executing the program in a non-parallel way. Then, we will show how to make these bag-identifiers invariant to parallelization. That is, how to make sure during the distributed execution, that bags with the same IDs are created as in the non-parallel execution, and the bags that each bag are computed from are also the same. Specifically, we will show how physical operator instances can determine during a distributed execution
\begin{enumerate}[itemsep=0em]

\item the ID of the output bag that they should compute next (\autoref{sec:choosing-out});

\item the IDs of the input bags that they should use to compute a particular output bag (\autoref{sec:choosing-input});

\item on which conditional output edge (see \autoref{sec:topology}) they should send a particular output bag (\autoref{sec:choosing-cond}).

\end{enumerate}

\subsubsection{Bag-identifiers with Execution Paths} \label{sec:bag-ids}
The bag-identifier encapsulates the identifier of the creating transformation and the execution path of the program up to the creation of the bag. The execution path is a \emph{walk} on the control flow graph, i.e., the sequence of basic blocks that the execution reached.

This is straightforward to keep track of during a non-pipelined, non-parallel execution. In this simple case, we execute the program on a single machine, with only one transformation being executed at a time, and fully materializing each bag before the next transformation starts consuming it. We keep track of the execution path in a global variable as a list of basic blocks. We append a new element to this list when a control flow instruction makes a decision about where the control flow should proceed. Each transformation can read this global variable to determine the ID of the bag that it is currently producing. As outlined at the beginning of \autoref{sec:cfl}, we will use this non-parallel execution as a specification of what bags the distributed execution should produce.

In the case of a distributed execution, the execution path is determined by the condition nodes, appending basic blocks when they evaluate conditions of if-statements or exit conditions of loops. Therefore, condition nodes broadcast these decisions to all the machines in the cluster, so that every physical instance of every operator knows how the execution path evolves. %

Note that there are no race conditions between condition nodes for who appends the next basic block (the execution path stays exactly the same as in a non-parallel execution). This is because each basic block has at most one condition node, and therefore it cannot happen that multiple condition nodes become active at the same time after a control flow decision.

There are basic blocks which have only one outgoing edge in the control flow graph, i.e., execution can proceed out from them in only one way (e.g., B and C in \autoref{fig:sync-example}). These basic blocks do not have any condition nodes, which would broadcast their successor block. Therefore, we make it the responsibility of a condition node that appends such a block to also append the next basic block. For example in \autoref{fig:sync-example}, when the condition node of the if-statement appends B to the execution path, it also appends D.

The execution path can grow arbitrarily large, which could conceivably introduce a bottleneck, if the coordination is implemented naively. For example, if the current execution path is repeatedly sent over the network as a list of basic blocks attached to every bag (as part of the bag ID), then we would require $\mathcal{O}(n^2)$ amount of network communication over the complete program execution, where $n$ is the length of the complete execution path. However, we can easily make sure that the actual implementation performs only $\mathcal{O}(1)$ work for every new basic block that is appended to the execution path, and therefore requires only $\mathcal{O}(n)$ work over the complete program execution:
\begin{itemize}
\item Condition nodes should broadcast only the currently added basic block;
\item When we send a bag ID over the network, it is enough to only send the length of the execution path in the ID, since every operator knows the current full list of basic blocks from the broadcasts of the condition nodes, and the execution paths in the bag IDs are always prefixes of the full list;
\item The procedures described in the next sections should not repeatedly scan the entire path, but incrementally keep track of any relevant information as the path evolves.
\end{itemize}

\subsubsection{Choosing Output Bags} \label{sec:choosing-out}

By watching how the execution path evolves, operators can choose the IDs of output bags to be computed in a straightforward way: when the path reaches the basic block of the operator, the operator starts to compute the bag whose bag ID contains the current path. For example, in Challenge 2 in \autoref{sec:chall-distr-coord}, this means that the physical operator instances of both $x_3$ and $y_3$ will choose to compute the output bag with path $ABD$ in its ID first, and then $ABDACD$.

Recall that we use the non-parallel execution described in \autoref{sec:bag-ids} as a specification of what bags the distributed execution should produce.
Since the execution path is the same during the distributed execution as in the non-parallel execution, this choice of output bags ensures that the sequence of bag IDs that each operator produces is indeed the same in the distributed execution as in the non-parallel execution.

\subsubsection{Choosing Input Bags} \label{sec:choosing-input}
When an operator $O_2$ decides to produce a particular output bag $g_2$ next, it also needs to choose input bags for it (see Challenge 2 in \autoref{sec:chall-distr-coord}). This choice is made independently for each logical input. 

In a non-parallel execution, the operator uses the latest bag that was written to the variable that the particular input refers to. We can mirror this behavior in the distributed execution by examining the execution path while keeping in mind the operator's basic block and the input's basic block.
More specifically, for a logical input $i$ of $O_2$, let $O_1$ be the operator whose output is connected to $i$, $b_1$ and $b_2$ be the basic blocks of $O_1$ and $O_2$, and $c$ be the execution path in the ID of $g_2$. To determine the bag ID that we use from among the bags coming from $i$ to compute $g_2$, we choose the execution path that is the longest such prefix of $c$ that ends with $b_1$. The reason behind this is that we need to use the latest bag computed by $O_1$.

We treat $\Phi$-nodes in a special way: For each particular output bag, a $\Phi$-node reads a bag from only one input. Therefore, we adapt the above procedure for choosing among the inputs by looking at the above-mentioned prefixes for each input, and choosing the longer one.

Note that sometimes we use a particular input bag for computing multiple output bags (see Challenge 1 in \autoref{sec:chall-distr-coord}). For this, we have to store that particular input bag, and thus also need to decide when to discard it: we watch as the execution path evolves and discard the input bag once the execution path reaches a basic block from which there is no such path on the control flow graph that reaches $b_2$ before $b_1$.

\subsubsection{Choosing Conditional Outputs} \label{sec:choosing-cond}
Operators look at how the execution path evolves after a particular output bag and send the bag on such conditional output edges (\autoref{sec:topology}) whose target is reached by the path before the next output bag is computed.

Specifically, let $O_1$ be an operator that is computing output bag $g$, $e$ be a conditional output edge of $O_1$, $O_2$ be the operator that is the target of $e$, $b_1$ be the basic block of $O_1$, $b_2$ be the basic block of $O_2$, and $c$ be the execution path of the ID of $g$. Note that the last element of $c$ is $b_1$.

$O_1$ should examine each new basic block appended to the execution path and send $g$ to $O_2$ when the path reaches $b_2$ for the first time after $c$ but before it reaches $b_1$ again. Note that this means that instances of $O_1$ can discard their partitions of $g$ once the execution path reaches such a basic block from which every path to $b_2$ on the control flow graph goes through $b_1$. (This can be understood by considering the non-SSA form of the program: the variable of $O_1$ will be overwritten before it is read by the variable of $O_2$.) Note that if $O_2$ is a $\Phi$-function, then we also need to consider the basic blocks of the other inputs of $O_2$.

\section{Optimization: Loop-Invariant Hoisting} \label{sec:loop-invariant-runtime}

We show here how to incorporate the well-known optimization described in \autoref{sec:static} into our system.

Normally, the bag-transformations implementing the interface shown in \autoref{sec:operator-interface} drop the state that they have built up during the computation of a specific output bag. However, to perform the optimization mentioned in \autoref{sec:static}, we extend the bag-transformation interface to allow the runtime to let the transformation implementations know when to keep their transformation-specific state that they build up for an input (e.g., so that a hash join can keep the hash table that it created for the build-side input and can use it across the computation of several output bags).

Assume, without loss of generality, that the first input of the transformation is the one that does not always change between output bags, and the second input changes for every output bag. We add a method named \emph{dropState} to the interface defined in \autoref{sec:operator-interface}. The runtime calls this method between computing two output bags to notify the transformation that the bag coming from the first input will change for the next output bag. In this case, the transformation should drop the state built-up for the first input.
Otherwise, the transformation implementation should assume that the first input will be the same bag as before, and that the runtime will not call \texttt{PushInElement}  for the elements of this bag.
For the example shown in \autoref{fig:nested-loops-example}, \emph{dropState} is called at every step of the outer loop, but not between steps of the inner loop.
For knowing when to call \emph{dropState}, the runtime can use the same mechanism that we described at the end of \autoref{sec:choosing-input} for keeping track of when to drop the buffered-up inputs.

\section{Implementation} \label{sec:impl}
We implemented the Labyrinth runtime on top of Flink.
For fault-tolerance, we can rely on the fault-tolerance of Flink, i.e., periodically creating consistent distributed snapshots of the states of all the operators in the dataflow job \cite{carbone2017state}.

We implemented the compilation procedure using the compiler infrastructure of the Emma language \cite{alexandrov2015implicit, alex2018representations}. Emma is a high-level DSL for scalable data analytics, embedded in Scala. It aims to reuse the language constructs of Scala as much as possible. For example, the user specifies the control flow by using while-loops, if-sta\-te\-ments, and mutable variables of the Scala language (as opposed to special-purpose functional constructs, as is the case in Flink, Naiad, and TensorFlow \cite{yu2018dynamic}).
Emma is able to extensively analyze and rewrite the user's program at the abstract syntax tree level, by employing a quotation-based embedding approach \cite{najd2016everything} (using Scala macros \cite{burmako2013scala}). As a first step of the compilation, Emma transforms the program into a typed SSA-like representation, which we can use as a starting point for the procedure described in \autoref{sec:ssa-to-single-dataflow}.

\section{Evaluation} \label{sec:eval}

In this section, we evaluate the performance benefits of compiling programs that have control flow to a single cyclic dataflow job, compared to launching separate dataflow jobs after every control flow step. First, we show microbenchmarks that verify that the overhead of Labyrinth's control flow coordination algorithm is orders of magnitude smaller than the overhead of launching a dataflow job. Then, we show that this difference can lead to significant scalability improvements in real workloads. Finally, we evaluate further optimizations that are enabled by having the entire program in a single dataflow job.

We ran the experiments on a cluster of 26 machines, connected by Gigabit Ethernet. The machines have two Intel Xeon E5620 CPUs (each with 4 physical cores), and 32 GB RAM.
We used Flink 1.6.0 and Spark 2.3.0. To reduce noise in the results, we performed each experiment at least three times, and report the median run time.

\subsection{Microbenchmarks for Control Flow Overhead} \label{sec:exp-sched-overhead}

\subsubsection{Scheduling Overhead}
We measured the scheduling overheads in Flink and Spark when executing minimal jobs that operate on only a minimal amount of data.
The jobs did not perform any disk IO, but created only a parallel collection distributed on the cluster. We set the parallelism of the collection to the recommended value for each system: the number of physical CPU cores for Flink, and two times that for Spark \cite{sparkparallelism}. %
We warmed up the JVMs by running a series of warm-up jobs before actually starting the timed job executions.

Of the 26 machines one was the master node and was also executing the client program, and a varying number of the other machines were used as workers. The systems had one worker process per machine.

\begin{figure}[t]
  \centering
  \begin{tikzpicture}
    \begin{axis}[
       xlabel={Number of worker machines},
       ylabel={Execution time (ms)},
       legend entries={\small Flink, \small Spark},
       legend style={at={(1.03,0.5)},anchor=west},
       width={6cm},
       height={3.7082039325cm},
       xtick={1, 9, 17, 25},
     ]
     \addplot [chartred, mark=*, mark options={scale=0.9},] table [x={Number-of-worker-machines}, y={Flink-median}] {scheduling.dat};
     \addplot [YellowOrange, mark=diamond*, mark options={scale=1}] table [x={Number-of-worker-machines}, y={Spark-median}] {scheduling.dat};
    \end{axis}
  \end{tikzpicture}
  \caption{Microbenchmark for the scheduling overhead.} %
  \label{fig:sched}
\end{figure}
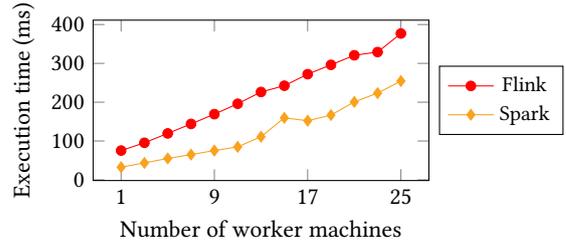

The size of the cluster has a significant impact on the scheduling overhead, therefore, we performed the experiment with a varying number of worker machines. We can see the run time of one job execution as a function of how many worker nodes are used in \autoref{fig:sched}. We can see that the execution time increased linearly, and it reached 254 ms and 376 ms in the 25 worker case with Spark and Flink, respectively.

\subsubsection{Iteration Step Overhead}

\begin{figure}[t]
  \centering
  \begin{tikzpicture}
    \begin{loglogaxis}[
       xlabel={Number of worker machines},
       ylabel={\small Time of a step (ms)},
       xticklabel={
         \pgfkeys{/pgf/fpu}
         \pgfmathparse{exp(\tick)}\pgfmathprintnumber\pgfmathresult
         \pgfkeys{/pgf/fpu=false}
       },
       xtick={1, 3, 5, 7, 9, 11, 13, 15, 17, 19, 21, 23, 25},
       xticklabels={1, 3, 5, 7, 9, , 13, , , 19, , , 25},
       extra y ticks={1, 10, 100, 1000},
       legend entries={
       \small Flink new jobs, %
       \small Spark new jobs,
       \small TensorFlow,
       \small Naiad,
       \small Flink in-dataflow,
       \small Labyrinth},
       legend style={at={(0.5,-0.35)},anchor=north,},
       legend style={/tikz/every even column/.append style={column sep=0.5cm}},
       legend columns=2,
       width={7.85cm},
       height={4.851cm},
     ]
     \addplot [chartred,] table [x={NumWorkers}, y={FlinkNewJobs}] {cf-microbenchmark.dat};
     \addplot [YellowOrange,] table [x={NumWorkers}, y={Spark}] {cf-microbenchmark.dat};
     \addplot [blue] table [x={NumWorkers}, y={TensorFlow}] {cf-microbenchmark.dat};
     \addplot [black] table [x={NumWorkers}, y={Naiad}] {cf-microbenchmark.dat};
     \addplot [brown] table [x={NumWorkers}, y={FlinkNative}] {cf-microbenchmark.dat};
     \addplot [chartgreen] table [x={NumWorkers}, y={Laby}] {cf-microbenchmark.dat};
    \end{loglogaxis}
  \end{tikzpicture}
  \caption{Log-log plot of the microbenchmark for the per-step overhead. Expressing all iteration steps in one dataflow job is approximately two orders of magnitude faster than launching new dataflow jobs for every iteration step (the top two lines).}
  \label{fig:step-overhead}
\end{figure}
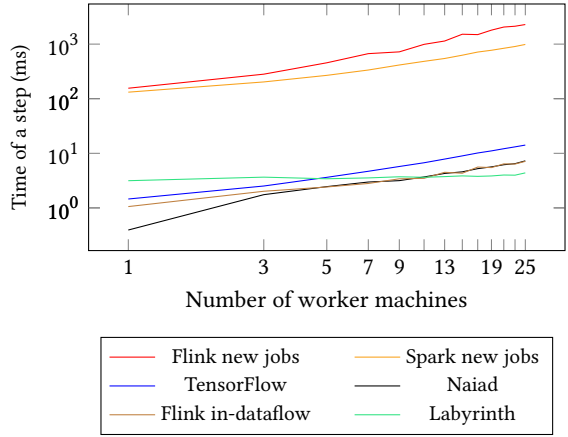

We performed a microbenchmark that measures the overhead of an iteration step in the different approaches of implementing control flow. This way, we (a) verified that the overhead of coordinating the distributed execution of control flow in Labyrinth is orders of magnitude smaller than the scheduling overhead, and (b) we found that Labyrinth's control flow coordination has a similar overhead to the approaches of Flink, Naiad, and TensorFlow.

The experiment program had a large number of steps and only a minimal amount of processing per step, and therefore the overhead of iteration steps dominated the execution time: %

\begin{samepage}
\begin{algorithmic}[0]
  \State i = 0
  \State bag = <new Bag with 200 elements>
  \Do
    \State i = i + 1
    \State bag = bag.map\{x => x + 1\}
  \DoWhile{i < numSteps%
  }
\end{algorithmic}
\end{samepage}

\autoref{fig:step-overhead} is a log-log plot of the run times of several different implementations of this program:

\textbf{Separate Dataflow Jobs.} The top two lines show when we launched new dataflow jobs for every step in Flink and Spark. The time per step here is similar to the scheduling overhead shown in the previous section, but a bit larger because of the additional operators. Note that in the Spark implementation we called \texttt{.cache} at every step and triggered the computation with a \texttt{.count} call.
Flink does not have an equivalent to \texttt{.cache}, so we collected the bag to the driver at each step.

\textbf{Flink and Naiad In-Dataflow Iterations.} Flink and Naiad offer higher-order functions for building an iteration into a dataflow, and thereby implementing an entire iteration inside a single dataflow job. The time per step is more than two orders of magnitude smaller here than with separate dataflow jobs.
In Flink, iteration steps are completely separated from each other by a global synchronization barrier. In contrast, Naiad's default behavior would be to continuously forward elements between iteration steps (independently between the parallel instances of the map operator). However, real loops typically have at least one pipeline-breaker in them (e.g., a grouped aggregation). For this reason, and also to make the comparison with Flink fair, in Naiad we artificially made the map operation into a pipeline-breaker, i.e., it has to receive a signal that all input elements belonging to an iteration step have arrived before sending its output to the next iteration step\footnote{The Naiad paper \cite{murray2013naiad} shows a somewhat similar experiment in Section 5.2, but with more than an order of magnitude better results. This difference can be due to many differences in the exact experimental setup: there was no data in their experiment, but only coordination; they executed Naiad on .Net on Windows, while we executed on Mono on Linux; differences in the network hardware; etc.}.

\textbf{TensorFlow.} Similarly to Naiad and Flink, TensorFlow also offers higher-order functions for building control flow into dataflows, and we were able to achieve a similar performance to Naiad and the Flink in-dataflow iterations\footnote{The TensorFlow paper \cite{yu2018dynamic} shows a similar experiment in Section 6.1, but with an order of magnitude better results. This difference is largely due to their experiment having only one operator instance per worker machine, while we created 8 (the number of CPU cores) to make the comparison with the other systems fair. (With only one instance per worker, the run time was $\sim$2ms for 25 machines in our setup.) %
}. Compared to Flink and Naiad, our TensorFlow implementation had the disadvantage that we needed to create an operator for the loop counter as well, since the TensorFlow control flow constructs do not have a built-in mechanism for having a fixed number of iteration steps.

\textbf{Labyrinth.} Finally, we have implemented the program in Labyrinth (the green line), which also builds the control flow into a single dataflow job. We can see that Labyrinth's control flow coordination algorithm has a similar overhead as the coordinations of Flink, Naiad, and TensorFlow. Note that to make the comparison fair, we artificially made the map operator into a pipeline breaker, similarly to the Naiad implementation. Note that, similarly to TensorFlow, there is a dataflow operator created for the loop counter variable.

\subsection{Strong Scaling with Real Workloads}

\subsubsection{Visit Count Example}

We performed an experiment with the Visit Count example program shown in \autoref{sec:example}, which compares visit counts of a set of webpages from consecutive days. For this experiment, we removed the join with the loop invariant dataset (Lines 1, 9). Note however that we concentrate on this part in \autoref{sec:eval-loop-invariant}. (We use only 100 iteration steps, instead of 365.)

The iteration in this program is not a fixpoint iteration, and therefore it cannot be implemented as a single dataflow job in Flink. We implemented the program in the Flink batch API (with separate dataflow jobs for each iteration step), in Spark, and in Labyrinth (with a single dataflow job).

We have seen in the previous experiments that the control flow overhead depends on the number of machines. Therefore, we test strong scaling here, i.e., we vary the number of machines, but keep the input data size fixed (19~GB in total, which is approximately 190~MB per day).

\begin{figure}[t]
  \centering
  \begin{tikzpicture}
    \begin{semilogyaxis}[
      xlabel={Number of worker machines},
      ylabel={\small Execution time (s)},
      legend entries={\small Spark,
      \small Flink,
      \small Labyrinth,
      \small Labyrinth pipelined,
      \small Single-threaded C++},
      legend style={at={(0.5,-0.35)},anchor=north},
      legend columns=2,
      width={7.85cm},
      height={4.851cm},
      extra y ticks={300, 700, 2000},
      extra y tick labels={
        \pgfkeys{/pgf/number format/sci} \pgfmathprintnumber{300},
        \pgfkeys{/pgf/number format/sci} \pgfmathprintnumber{700},
        \pgfkeys{/pgf/number format/sci} \pgfmathprintnumber{2000}
      },
      xmin=-1,xmax=27,
      ]
      \addplot [YellowOrange, mark=diamond*, mark options={scale=1.1},] table [x={numMachines}, y={Spark}] {scaling-click-count.dat};
      \addplot [chartred, mark=*, mark options={scale=0.9},] table [x={numMachines}, y={nolaby}] {scaling-click-count.dat};
      \addplot [blue, mark=triangle*, mark options={scale=1},] table [x={numMachines}, y={Laby-barrier}] {scaling-click-count.dat};
      \addplot [chartgreen, mark=square*, mark options={scale=0.9},] table [x={numMachines}, y={Laby-nobarrier}] {scaling-click-count.dat};
      \addplot [cyan, dashed, domain=-1:27] {723.8};
    \end{semilogyaxis}
  \end{tikzpicture}
  \caption{Strong scaling for the Visit Count example (without the join with the loop-invariant dataset).}
  \label{fig:scaling-click-count}
\end{figure}
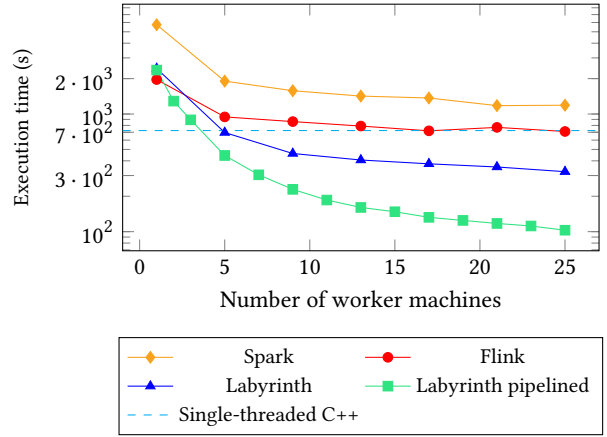

Looking at Flink and Labyrinth in \autoref{fig:scaling-click-count}, we can see that with only a few machines the execution times were similar. However, when we increased the number of machines, Flink fell behind by approximately a factor of two, as the scheduling overhead became more significant. Note that the per-step overhead of Flink was larger than in the microbenchmark, since there are more operators inside the loop body in this real workload than in the microbenchmark, and thus Flink has to schedule more operators.

The pipelined version of Labyrinth (green), which we discuss in \autoref{sec:eval-pipelining}, was a further three times faster with 25 machines.

Spark had a similar scaling behavior to Flink, because of similarly launching separate dataflow jobs (but was slower than Flink with all machine counts).

As McSherry et al.\ \cite{mcsherry2015scalability} pointed out, it can be interesting to compare the performance of distributed data analytics frameworks to single-threaded implementations that do not use any such frameworks. Therefore, we wrote a single-threaded implementation in C++, using only the STL. Both the reduceByKey and the join are implemented using sorting. As we can see in \autoref{fig:scaling-click-count}, Flink and Spark are not able to outperform the single-threaded implementation even with 25 machines, while (even the non-pipelined version of) Labyrinth achieves this with only 5 machines, and surpasses it by a factor of about 2.2 at 25 machines.

\subsubsection{PageRank}

We performed an experiment with a modified version of the Visit Count example in \autoref{fig:example-src-imp}: instead of page visits we read page transitions (pairs of pages), and instead of counting page visits we computed the PageRank \cite{page1999pagerank} of the graph of page transitions. That is, we replaced Line 10 by an inner loop, which computes the PageRank.

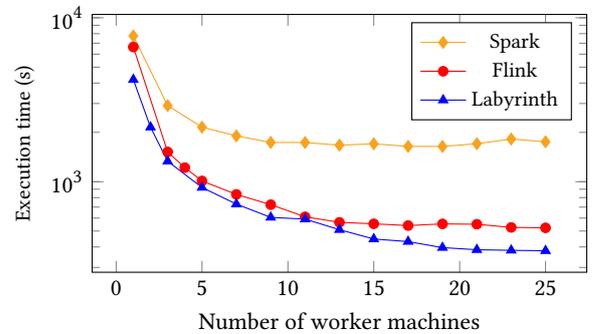
\begin{figure}[t]
  \centering
  \begin{tikzpicture}
    \begin{semilogyaxis}[
       xlabel={Number of worker machines},
       ylabel={\small Execution time (s)},
       legend entries={\small Spark,
       \small Flink,
       \small Labyrinth},
       legend pos={north east},
        width={8.09016994375cm},
        height={5cm},
     ]
     \addplot [YellowOrange, mark=diamond*, mark options={scale=1.1},] table [x={numMachines}, y={Spark}] {PageRank.dat};
     \addplot [chartred, mark=*, mark options={scale=0.9},] table [x={numMachines}, y={NoLaby}] {PageRank.dat};
     \addplot [blue, mark=triangle*, mark options={scale=1},] table [x={numMachines}, y={Laby}] {PageRank.dat};
    \end{semilogyaxis}
  \end{tikzpicture}
  \caption{Strong scaling with PageRank. (The join with the loop-invariant dataset is not present here.)} %
  \label{fig:scaling-PageRank}
\end{figure}

We plotted the run time as a function of the number of machines in \autoref{fig:scaling-PageRank}. PageRank can be expressed as a fixpoint iteration, and therefore the inner loop can be expressed in Flink as a single dataflow job. We can see that this results in Flink having a very similar scaling behavior as Labyrinth. Note that even though Flink has the scheduling overhead for each step of the outer loop (which cannot be expressed as a fixpoint iteration), this was not a significant part of the run time.

In the case of Spark, however, every step of both the inner and outer loops have to be expressed as a separate dataflow job, which results in a worse scaling behavior. With just one machine, the run time of Spark was within a factor of two from Labyrinth. However, with a large number of machines, Spark was about 4.62 times slower than Labyrinth. We can also see that Spark was not getting faster when we increase the number of machines beyond 9, while Labyrinth was getting faster until about 20 machines.

\subsection{\fontsize{11.9pt}{0pt}\selectfont Optimization: Loop Pipelining} \label{sec:eval-pipelining}

Systems that start new dataflow jobs for every iteration step, typically wait for an iteration step to be fully finished before starting the next step. In Labyrinth, all iteration steps are processed by the same dataflow job, which makes it possible that some operators start processing the next iteration step while others are still processing the previous one. For example, in the Visit Count example program (\autoref{fig:example-src-imp}), lines 5--11 do not have dependencies on computations of previous iteration steps. This means that they can proceed with subsequent iteration steps independently from the operators in Lines 14--17.

In \autoref{fig:scaling-click-count}, the green line shows the times when this optimization is enabled (this is the default execution mode of Labyrinth, as the control flow coordination algorithm described in \autoref{sec:distr-coord} already provides this property without further extension).
With 25 machines, it was about 3 times faster than the execution with barriers (blue line). Also note that it had close to ideal scaling behavior, as it was about 23 times faster with 25 machines than with one machine.

\subsection{\fontsize{11.9pt}{0pt}\selectfont Optimization: Loop-Invariant Hoisting} \label{sec:eval-loop-invariant}

\begin{figure}[t]
  \centering
  \begin{tikzpicture}
    \begin{loglogaxis}[
        xlabel={Data size scale},
        ylabel={\small Execution time (s)},
        legend style={cells={align=left}}, %
        legend entries={\small Flink, \small Laby-no-reuse, \small Laby-reuse},
        legend style={at={(1.03,0.5)},anchor=west},
        width={5.6cm},
        height={3.461cm},
        xtick={1, 2, 4, 8, 16, 32},
        xticklabel={
            \pgfkeys{/pgf/fpu}
            \pgfmathparse{exp(\tick)}\pgfmathprintnumber\pgfmathresult
            \pgfkeys{/pgf/fpu=false}
        },
     ]
     \addplot [chartred, mark=*,] table [x={Size}, y={Flink}] {click-count.dat};
     \addplot [blue, mark=triangle*, mark options={scale=1},] table [x={Size}, y={Laby-no-reuse}] {click-count.dat};
     \addplot [chartgreen, mark=square*,] table [x={Size}, y={Laby}] {click-count.dat};
    \end{loglogaxis}
  \end{tikzpicture}
  \caption{The run time of the Visit Count example with varying data sizes. A scale of 1 means a 10MB dataset per day and a 251MB \emph{pageAttributes} dataset. For ``Laby-no-reuse'' we turn off the optimization of reusing the build-side of the hash join with loop-invariant data.}
  \label{fig:click-count-plot}
\end{figure}
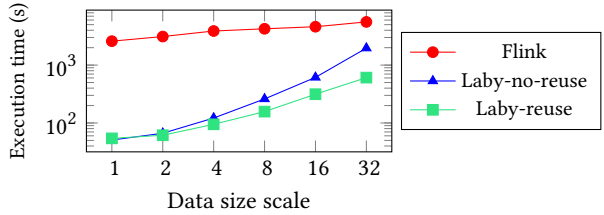

Here, we evaluate the optimization of \autoref{sec:loop-invariant-runtime}. We use the Visit Count example shown in \autoref{fig:example-src-comparison}, and concentrate on the join with the loop-invariant dataset (Line 9). \autoref{fig:click-count-plot} is a log-log plot of the execution time of the different implementations when the size of the input data is varied. (The number of worker machines is fixed at 25.)

The blue line shows the Labyrinth execution where we have turned off the optimization of reusing the hash table of the join, while the green line shows the execution with the optimization turned on. For the largest data size, this optimization gave a factor of three speedup. For the two smallest data sizes its effect was negligible, which is due to the iteration step overhead being the dominant factor in the execution time even with Labyrinth.

\section{Related Work} 
\label{sec:related_work}

The dataflow model of computing has a long history \cite{whiting1994history}. Arvind et al. \cite{culler1986dataflow} include control flow into dataflow graphs through the \emph{switch} and \emph{merge} primitives (which TensorFlow recently adopted \cite{yu2018dynamic}). Gamma \cite{dewitt1986high} is a database machine built on the dataflow model. In the following paragraphs, we concentrate on systems for large-scale, distributed data processing.

\textbf{Adding Iterations to MapReduce.} The CGL-MapReduce \cite{ekanayake2008mapreduce}, HaLoop~\cite{bu2010haloop}, Twister \cite{ekanayake2010twister}, and iMapReduce \cite{zhang2012imapreduce} systems extend MapReduce \cite{dean2004mapreduce} to provide better support for iterations. The programming model of these systems is based on repeatedly executing MapReduce programs rather than building more complex programs using a collection-based API.

\textbf{Spark} \cite{zaharia2010spark} provides better support for iteration than MapReduce in that it allows for keeping the data in memory between iteration steps. However, the scheduling overhead is still present at every iteration step. Furthermore, Spark cannot automatically hoist loop-invariant computations out of loops.
Note that for a hash join whose only one input is static (e.g., in \autoref{fig:example-src-comparison}) even manually performing this optimization is not possible, since this would require breaking up the hash join into two operators where one of them would have to look in the other's internal hash table, which is not supported by the RDD interface.

\textbf{In-Dataflow Control Flow.} Flink \cite{ewen2012spinning}, Naiad \cite{murray2013naiad}, and TensorFlow \cite{yu2018dynamic} allow for building control flow into dataflows, but only through functional APIs (see \autoref{sec:func-api}). On the other hand, Labyrinth allows the user to write control flow in the familiar imperative style. Also note that the control flow representations in these systems do not map directly to SSA, which would make the compilation from imperative control flow more involved. Nevertheless, there are open-source efforts to compile imperative control flow written in Swift \cite{swiftcf} and Python \cite{autograph} to TensorFlow's low-level control flow API. %

\textbf{Differences to TensorFlow's Coordination Algorithm.} \\
Operators potentially execute multiple times in the presence of control flow. TensorFlow uses the current iteration numbers of all enclosing loops to identify these executions, while Labyrinth uses the execution path. This design choice results naturally from using SSA end-to-end in Labyrinth, as the execution path gives us a lot of useful information for various aspects of our coordination (see sections \ref{sec:bag-ids}--\ref{sec:choosing-cond}). 
Also note that Labyrinth's algorithm avoids introducing the \texttt{is\_dead} signal of TensorFlow.

\textbf{Dynamic Dataflows.} The CIEL system \cite{murray2011ciel} supports dynamic dataflows in the sense that a running job can spawn new tasks, essentially extending the dataflow. By this mechanism, the system supports the implementation of control flow constructs. The Aura dataflow engine \cite{herb2016aura} supports control flow by scheduling dataflow-fragments belonging to individual basic blocks at the time when control flow enters the corresponding basic block.
In the above systems, scheduling overhead occurs between every iteration step, which we avoid in Labyrinth by constructing a single dataflow job for the entire iteration.

\textbf{Faster Scheduling.} The systems mentioned above employ a centralized scheduler, which places a fundamental limit on the scheduling speed. The Sparrow scheduler \cite{ousterhout2013sparrow} achieves faster scheduling by distributing the scheduling itself, thereby reducing the scheduling overhead. %

\textbf{Loop-Invariant Hoisting} is a well-known optimization in the context of distributed data analytics systems \cite{ewen2012spinning, murray2013naiad, bu2010haloop, zhang2012imapreduce, ekanayake2010twister}. Labyrinth differs from the referred systems in that it is able to perform this optimization even when the user expressed the program using imperative control flow constructs.

\textbf{SystemML} \cite{boehm2016systemml} is also able \cite{systemmlloopinvarianthoisting} to perform loop-invariant hoisting in the presence of imperative control flow. However, it is not able to perform this optimization on a binary operator which has only one static input (e.g., the hash join that we used in \autoref{sec:eval-loop-invariant}), since it compiles to Spark.
As mentioned in \autoref{sec:compilation-assumptions}, SystemML's language could be compiled to Labyrinth dataflows.
Also note that SystemML makes a decision whether to execute a computation locally or in the cluster based on data sizes, which makes the overhead of launching a dataflow job disappear for small enough data sizes.

\section{Conclusion}
\label{sec:conclusion}

End-to-end data analysis often requires complex control flow, while efficient execution requires that the whole program is expressed as a single dataflow job. However, existing systems can represent control flow inside a single dataflow job only if the user expresses control flow using functional APIs, which are hard to use. In this paper, we propose Labyrinth, which allows users to express control flow by easy-to-use imperative constructs, and still executes these programs efficiently as a single dataflow job. To support general control flow, we rely on SSA both in our compilation procedure and in our algorithm for coordinating the distributed execution of control flow.
Our experimental evaluation shows that Labyrinth reduces the overhead of iteration steps by orders of magnitude compared to launching new dataflow jobs at every step, %
and allows for optimizations that result in significant speedups.

\section*{Acknowledgments}
{
\small
We would like to thank Alexander Alexandrov for pointing our attention to SSA and for helpful suggestions on improving the writing, and G\'abor Hermann, Georgi Krastev, Clemens Lutz, and Felix Neutatz for insightful comments on drafts of the paper.
This work was partly funded by the EU project E2Data (780245), and the German Ministry for Education and Research as BBDC (01IS14013A).
\renewcommand{\baselinestretch}{1}
}

\balance

{\footnotesize \bibliographystyle{acm}
\bibliography{db}}

\begin{thebibliography}{10}

\bibitem{abadi2016tensorflow}
{\sc Abadi, M., Barham, P., Chen, J., Chen, Z., Davis, A., Dean, J., Devin, M.,
  Ghemawat, S., Irving, G., Isard, M., et~al.}
\newblock Tensor{F}low: A system for large-scale machine learning.
\newblock In {\em OSDI\/} (2016), vol.~16, pp.~265--283.

\bibitem{aho2007compilers}
{\sc Aho, A.~V., Sethi, R., and Ullman, J.~D.}
\newblock {\em Compilers: principles, techniques, and tools}, vol.~2.
\newblock Addison-wesley Reading, 2007.

\bibitem{alexandrov2014stratosphere}
{\sc Alexandrov, A., Bergmann, R., Ewen, S., Freytag, J.-C., Hueske, F., Heise,
  A., Kao, O., Leich, M., Leser, U., Markl, V., et~al.}
\newblock The {S}tratosphere platform for big data analytics.
\newblock {\em The VLDB Journal 23}, 6 (2014), 939--964.

\bibitem{alex2018representations}
{\sc Alexandrov, A., Krastev, G., and Markl, V.}
\newblock Representations and optimizations for embedded parallel dataflow
  languages.
\newblock {\em ACM Trans. Datab. Syst.\/}, (to appear).

\bibitem{alexandrov2015implicit}
{\sc Alexandrov, A., Kunft, A., Katsifodimos, A., Sch{\"u}ler, F., Thamsen, L.,
  Kao, O., Herb, T., and Markl, V.}
\newblock Implicit parallelism through deep language embedding.
\newblock In {\em Proceedings of the 2015 ACM SIGMOD International Conference
  on Management of Data\/} (2015), ACM, pp.~47--61.

\bibitem{culler1986dataflow}
{\sc Arvind, and Culler, D.~E.}
\newblock Dataflow architectures.
\newblock {\em Annual review of computer science 1}, 1 (1986), 225--253.

\bibitem{boehm2016systemml}
{\sc Boehm, M., Dusenberry, M.~W., Eriksson, D., Evfimievski, A.~V., Manshadi,
  F.~M., Pansare, N., Reinwald, B., Reiss, F.~R., Sen, P., Surve, A.~C.,
  et~al.}
\newblock System{ML}: Declarative machine learning on {S}park.
\newblock {\em Proceedings of the VLDB Endowment 9}, 13 (2016), 1425--1436.

\bibitem{bu2010haloop}
{\sc Bu, Y., Howe, B., Balazinska, M., and Ernst, M.~D.}
\newblock Ha{L}oop: efficient iterative data processing on large clusters.
\newblock {\em Proceedings of the VLDB Endowment 3}, 1-2 (2010), 285--296.

\bibitem{burmako2013scala}
{\sc Burmako, E.}
\newblock Scala macros: let our powers combine!: on how rich syntax and static
  types work with metaprogramming.
\newblock In {\em Proceedings of the 4th Workshop on Scala\/} (2013), ACM.

\bibitem{carbone2017state}
{\sc Carbone, P., Ewen, S., F{\'o}ra, {\relax Gy}., Haridi, S., Richter, S.,
  and Tzoumas, K.}
\newblock State management in {A}pache {F}link{\textregistered}: consistent
  stateful distributed stream processing.
\newblock {\em Proceedings of the VLDB Endowment 10}, 12 (2017), 1718--1729.

\bibitem{chambers2010flumejava}
{\sc Chambers, C., Raniwala, A., Perry, F., Adams, S., Henry, R.~R., Bradshaw,
  R., and Weizenbaum, N.}
\newblock Flume{J}ava: easy, efficient data-parallel pipelines.
\newblock In {\em ACM Sigplan Notices\/} (2010), vol.~45, ACM, pp.~363--375.

\bibitem{dean2004mapreduce}
{\sc Dean, J., and Ghemawat, S.}
\newblock Map{R}educe: Simplified data processing on large clusters.
\newblock {\em OSDI\/} (2004).

\bibitem{dewitt1986high}
{\sc DeWitt, D.~J., Gerber, R., Graefe, G., Heytens, M., Kumar, K., and
  Muralikrishna, M.}
\newblock {\em A High Performance Dataflow Database Machine}.
\newblock Computer Science Department, University of Wisconsin, 1986.

\bibitem{ekanayake2010twister}
{\sc Ekanayake, J., Li, H., Zhang, B., Gunarathne, T., Bae, S.-H., Qiu, J., and
  Fox, G.}
\newblock Twister: a runtime for iterative {M}ap{R}educe.
\newblock In {\em Proceedings of the 19th ACM international symposium on high
  performance distributed computing\/} (2010), ACM, pp.~810--818.

\bibitem{ekanayake2008mapreduce}
{\sc Ekanayake, J., Pallickara, S., and Fox, G.}
\newblock Map{R}educe for data intensive scientific analyses.
\newblock In {\em eScience, 2008. eScience'08. IEEE Fourth International
  Conference on\/} (2008), IEEE, pp.~277--284.

\bibitem{ewen2012spinning}
{\sc Ewen, S., Tzoumas, K., Kaufmann, M., and Markl, V.}
\newblock Spinning fast iterative data flows.
\newblock {\em Proceedings of the VLDB Endowment 5}, 11 (2012), 1268--1279.

\bibitem{ghoting2011systemml}
{\sc Ghoting, A., Krishnamurthy, R., Pednault, E., Reinwald, B., Sindhwani, V.,
  Tatikonda, S., Tian, Y., and Vaithyanathan, S.}
\newblock System{ML}: Declarative machine learning on {M}ap{R}educe.
\newblock In {\em 2011 IEEE 27th International Conference on Data
  Engineering\/} (2011), IEEE, pp.~231--242.

\bibitem{herb2016aura}
{\sc Herb, T., Thamsen, L., Renner, T., and Kao, O.}
\newblock Aura: A flexible dataflow engine for scalable data processing.
\newblock In {\em Tools for High Performance Computing 2015}. Springer, 2016,
  pp.~117--126.

\bibitem{isard2007dryad}
{\sc Isard, M., Budiu, M., Yu, Y., Birrell, A., and Fetterly, D.}
\newblock Dryad: distributed data-parallel programs from sequential building
  blocks.
\newblock In {\em ACM SIGOPS Operating Systems Review\/} (2007), vol.~41, ACM,
  pp.~59--72.

\bibitem{mcsherry2015scalability}
{\sc McSherry, F., Isard, M., and Murray, D.~G.}
\newblock Scalability! {B}ut at what {COST}?
\newblock In {\em HotOS\/} (2015), vol.~15, Citeseer, pp.~14--14.

\bibitem{murray2013naiad}
{\sc Murray, D.~G., McSherry, F., Isaacs, R., Isard, M., Barham, P., and Abadi,
  M.}
\newblock Naiad: a timely dataflow system.
\newblock In {\em Proceedings of the Twenty-Fourth ACM Symposium on Operating
  Systems Principles\/} (2013), ACM, pp.~439--455.

\bibitem{murray2011ciel}
{\sc Murray, D.~G., Schwarzkopf, M., Smowton, C., Smith, S., Madhavapeddy, A.,
  and Hand, S.}
\newblock {CIEL}: a universal execution engine for distributed data-flow
  computing.
\newblock In {\em Proc. 8th ACM/USENIX Symposium on Networked Systems Design
  and Implementation\/} (2011), pp.~113--126.

\bibitem{najd2016everything}
{\sc Najd, S., Lindley, S., Svenningsson, J., and Wadler, P.}
\newblock Everything old is new again: quoted domain-specific languages.
\newblock In {\em Proceedings of the 2016 ACM SIGPLAN Workshop on Partial
  Evaluation and Program Manipulation\/} (2016), ACM, pp.~25--36.

\bibitem{ousterhout2013sparrow}
{\sc Ousterhout, K., Wendell, P., Zaharia, M., and Stoica, I.}
\newblock Sparrow: distributed, low latency scheduling.
\newblock In {\em Proceedings of the Twenty-Fourth ACM Symposium on Operating
  Systems Principles\/} (2013), ACM, pp.~69--84.

\bibitem{page1999pagerank}
{\sc Page, L., Brin, S., Motwani, R., and Winograd, T.}
\newblock The {P}age{R}ank citation ranking: Bringing order to the web.
\newblock Tech. rep., Stanford InfoLab, 1999.

\bibitem{rastello2016ssa}
{\sc Rastello, F.}
\newblock {\em {SSA}-based Compiler Design}.
\newblock Springer Publishing Company, Incorporated, 2016.

\bibitem{rompf2012lightweight}
{\sc Rompf, T., and Odersky, M.}
\newblock Lightweight modular staging: a pragmatic approach to runtime code
  generation and compiled {DSL}s.
\newblock {\em Communications of the ACM 55}, 6 (2012), 121--130.

\bibitem{whiting1994history}
{\sc Whiting, P.~G., and Pascoe, R.~S.}
\newblock A history of data-flow languages.
\newblock {\em IEEE Annals of the History of Computing 16}, 4 (1994), 38--59.

\bibitem{yu2018dynamic}
{\sc Yu, Y., Abadi, M., Barham, P., Brevdo, E., Burrows, M., Davis, A., Dean,
  J., Ghemawat, S., Harley, T., Hawkins, P., et~al.}
\newblock Dynamic control flow in large-scale machine learning.
\newblock In {\em Proceedings of the Thirteenth EuroSys Conference\/} (2018),
  ACM, p.~18.

\bibitem{yu2008dryadlinq}
{\sc Yu, Y., Isard, M., Fetterly, D., Budiu, M., Erlingsson, {\'U}., Gunda,
  P.~K., and Currey, J.}
\newblock Dryad{LINQ}: A system for general-purpose distributed data-parallel
  computing using a high-level language.
\newblock In {\em OSDI\/} (2008), vol.~8, pp.~1--14.

\bibitem{zaharia2012resilient}
{\sc Zaharia, M., Chowdhury, M., Das, T., Dave, A., Ma, J., McCauley, M.,
  Franklin, M.~J., Shenker, S., and Stoica, I.}
\newblock Resilient distributed datasets: A fault-tolerant abstraction for
  in-memory cluster computing.
\newblock In {\em Proceedings of the 9th USENIX conference on Networked Systems
  Design and Implementation\/} (2012), USENIX Association.

\bibitem{zaharia2010spark}
{\sc Zaharia, M., Chowdhury, M., Franklin, M.~J., Shenker, S., and Stoica, I.}
\newblock Spark: cluster computing with working sets.
\newblock {\em HotCloud 10\/} (2010).

\bibitem{zhang2012imapreduce}
{\sc Zhang, Y., Gao, Q., Gao, L., and Wang, C.}
\newblock i{M}ap{R}educe: A distributed computing framework for iterative
  computation.
\newblock {\em Journal of Grid Computing 10}, 1 (2012), 47--68.

\bibitem{sparkparallelism}
{Spark Documentation on Performance Tuning}.
\newblock
  \url{https://spark.apache.org/docs/latest/tuning.html#level-of-parallelism},
  2018.
\newblock [Online; accessed 31-August-2018].

\bibitem{autograph}
{AutoGraph}.
\newblock \url{https://www.tensorflow.org/guide/autograph}, 2018.
\newblock [Online; accessed 12-September-2018].

\bibitem{swiftcf}
{Compiling Control Flow in Swift}.
\newblock \\
  \url{https://github.com/tensorflow/swift/blob/master/docs/GraphProgramExtraction.md#adding-intraprocedural-within-a-function-control-flow},
  2018.
\newblock [Online; accessed 12-September-2018].

\bibitem{systemmlloopinvarianthoisting}
{Loop-Invariant Hoisting in SystemML}.
\newblock
  \url{https://github.com/apache/systemml/blob/2cf78819d87a12afbda7ccf08ffa0d894e3207c7/src/main/java/org/apache/sysml/hops/rewrite/RewriteHoistLoopInvariantOperations.java},
  2018.
\newblock [Online; accessed 12-September-2018].

\end{thebibliography}

\end{document}